\documentclass[11pt]{article}
\usepackage{graphicx}
\usepackage{txfonts}
\usepackage{natbib}
\usepackage{color}
\usepackage[top=1.0in, bottom=1.0in, left=1.0in, right=1.0in]{geometry}
\newcommand{\mgc}[1]{{\color{black} #1}}

\begin{document} 

\title{Reduced drift-kinetics with thermal velocity distribution across magnetic field}

\author{Mykola Gordovskyy\thanks{e-mail: mykola.gordovskyy [AT] manchester.ac.uk} $\,$ and Philippa Browning\\
Jodrell Bank Centre for Astrophysics, University of Manchester, Manchester M13 9PL, UK}

\date{}

\maketitle

%
%________________________________________________________________

{\bf Abstract.} The goal of this study is to develop an approximate self-consistent description of particle motion in strongly magnetised solar corona. We derive a set of reduced drift-kinetic equations based on the assumption that the gyro-velocity distribution is Maxwellian. The equations are tested using simple 1D models. 

\section{Introduction}

Generally, solving even the 2D drift-kinetic problem would require at least a 4-dimensional phase space (2D2V). This is practically difficult: taking into account a number of processors normally available (up to $\sim$100) and a `reasonable' wall-time (up to 3-5 days), it would be possible to consider a domain with up to $\sim$10$^7$-10$^8$ grid points. Realistically, the phase-space can have 3 dimensions with a reasonable resolution (e.g. 2D1V), or 4 dimensions (e.g. 2D2V) with very scarce resolution along at least one of dimensions. Hence, full kinetic (or drift-kinetic) treatment would be too numerically expensive in case of 2D or 3D geometry.

Various analytical studies and numerical simulations \cite[e.g.][]{gore10,gobr11} show that particle acceleration by quasi-stationary electric fields in the solar corona affects the parallel component of particle velocity $v_{||}= \vec{v} \cdot \vec{b}$ (where $\vec{b} = \vec{B}/B$ is the magnetic field direction), while the gyro-velocities $v_g$ normally remain nearly thermal. This is natural, taking into account that in the corona magnetic field curvature is small compared to particle Larmor radii and collisional times are longer than acceleration times, i.e. there are no strong scattering mechanisms, and accelerated particles are expected to remain collimated along magnetic field. 

Kinetic description of particles with small pitch-angles (i.e. $v_{||} \gg v_g$) has been discussed in number of papers in the laboratory plasma context \cite[e.g.][]{pfee15}. Generally, an approximation of zero Larmor velocities, at least, for energetic particles, would substantially simplify the kinetic equations, removing a dimension from the phase space. However, in context of large-scale particle kinetics in the solar corona, Larmor radii, even small, may play an important role in some cases, for instance, during particle mirroring from the strongly converging magnetic field at the bottom of the corona. Therefore, it might be more realistic to assume that particle gyro-velocities are, generally, non-zero, but their distributions remain Maxwellian. 

Assuming the gyro-velocities always have Maxwellian distribution, one could reduce the phase space by looking for a distribution function $F(\vec{r},v_{||}; t)$ and the perpendicular temperature $\tau(\vec{r},v_{||}; t)$ instead of the distribution function $f(\vec{r},\vec{v}; t)$. Here, the perpendicular temperature defines the width of the Maxwellian distribution of gyro-velocities for a given particle specii with given parallel velocity at a given location. Hence, using this formalism the problem can be reduced to calculation of $F(x,y,v_{||}; t)$ and $\tau(x,y,v_{||}; t)$ in 2D case, and to calculation of $F(x,y,z,v_{||}; t)$ and $\tau(x,y,z,v_{||}; t)$ in 3D case. Below this formalism will be called `reduced kinetics'.

We derive a set of reduced drift kinetic equations by averaging the Larmor gyration velocity at each location of the phase space ($\vec{r},v_{||}$). The averaging is done assuming that the distribution in respect of gyro-velocity always remains Maxwellian. The resulting equations are required to conserve the particle number and energy. 

\section{Reduced equation derivation}

\subsection{Full kinetic equation}

Consider a kinetic equation in the following form:

\begin{equation}
\hat{L}f = \frac {\partial f}{\partial t} + \left( \vec{u} + v_{||} \vec{b} \right) \frac {\partial f}{\partial \vec{r}} + \frac{dv_{||}}{dt} \frac {\partial f}{\partial v_{||}} + \frac{dv_g^2}{dt} \frac {\partial f}{\partial v_g^2} = 0,
\label{eq-fkin}
\end{equation}
where the distribution function is $f=f(\vec{r},v_{||},v_g^2; t)$, $\vec{u}$ is the guiding centre drift velocity, $\vec{b}=\vec{B}/B$ is the magnetic field direction vector. This equation is adopted from a standard form of drift-kinetic equation for particles with non-zero magnetic moments \cite[see e.g.]{kuls83}.

The drift velocity consists of the following terms:
\begin{equation}
\vec{u} = \vec{u}^* + \vec{u}_{\nabla B}, 
\end{equation}
where $\vec{u}^* = \vec{u}_E + \vec{u}_C$ is the sum of ExB and curvature drifts
\[
\vec{u}_{E} = \frac{\vec{E}\times \vec{B}}{B^2},
\]
\[
\vec{u}_{C} = \frac{m v_{||}^2}{qB} [\vec{b} \times (\vec{b} \cdot \vec{\nabla}) \vec{b}],
\]
which don't depend on the gyrovelocity, and 
\[
\vec{u}_{\nabla B} = \frac mq \frac{\vec{B} \times \vec{\nabla}B}{2 B^2} v_g^2.
\]

Parallel velocity can be affected by the parallel electric field and by magnetic field gradient along magnetic field lines:
\begin{equation}
\label{eq-apara}
\frac{dv_{||}}{dt} = \frac qm \vec{\mathcal{E}}\cdot \vec{b} - v_g^2 \frac {\vec{\nabla} B \cdot \vec{b}}{2 B},
\end{equation} 
where $\vec{\mathcal{E}}$ is electric field, and $B$ and $\vec{b}$ are the absolute value and direction ($\vec{b}=\vec{B}/B$) of the magnetic field.

Variation of gyro-velocity can be derived from the magnetic moment conservation $v_g^2/B = \mathrm{const}$. Differentiating this in respect of time, and substituting 
$\frac{dB}{dt} = \frac{\partial B}{\partial t} + (v_{||} \vec{b} + \vec{u} ) \cdot \vec{\nabla} B$ gives
\begin{equation}
\label{eq-agyro}
\frac{d v_g^2}{dt} = \frac {\vec{\nabla} B \cdot \vec{b}}B v_g^2 v_{||} + \frac {\vec{\nabla} B \cdot \vec{u}}B v_g^2.
\end{equation}
The first term in the RHS corresponds to the magnetic mirroring effect, so that $2 v_{||} \frac{dv_{||}}{dt}+\frac{dv_g^2}{dt} = 2 v_{||} \frac{dv_{||}}{dt} +\frac {\vec{\nabla} B \cdot \vec{b}}B v_g^2 v_{||}=0$.
(We ignore terms containing $\frac{\partial E}{\partial t}$ and $\frac{\partial B}{\partial t}$, assuming that field variation timescale is much longer than $1/\omega_g$ of considered particle species.)

\subsection{Integrated distribution function and average gyro-velocity}

In terms of the parallel velocity $v_{||}$ and squared gyro-velocity $v_g^2$ the thermal distrbution with total specific energy $\mathcal{E}_{th}(\vec{r})= \int \limits_{-\infty}^{+\infty} \int \limits_{0}^{+\infty} (v_{||}^2 + v_g^2) F dv_{g}^2 dv_{||}$ and total particle number $\mathcal{N}_{th}(\vec{r}) = \int \limits_{-\infty}^{+\infty} \int \limits_{0}^{+\infty}F dv_{g}^2 dv_{||}$ is 

\begin{equation}
\label{eq-dfth}
f_{th}(v_{||},v_g^2) = f_{th\,0} \exp\left(-\frac{v_{||}^2 + v_g^2}{\tau_{th}}\right),
\end{equation}
where $f_{th\,0} = \frac {\mathcal{N}_{th}}{\sqrt \pi  \tau_{th}^{3/2}}$ and the "equivalent temperature" $\tau_{th}= \frac 23 \frac {\mathcal{E}_{th}}{\mathcal{N}_{th}} $.

We assume that the distribution function can be written in the following form:

\mgc{
\begin{equation}
\label{eq-fmaxw}
f(\vec{r},v_{||},v_g^2; t) = P(\vec{r},v_{||}; t) \frac{v_0^2}{\tau} \exp\left(-\frac{v_g^2}{\tau(\vec{r},v_{||}; t)}\right) = P(\vec{r},v_{||}; t) \mathcal{S}(v_g^2, \tau(\vec{r},v_{||}; t)) ,
\end{equation}
}
where $v_g$ is gyro-velocity and $v_0$ is some characteristic constant velocity.

Let us introduce a new distribution function integrated in respect of $v_g^2$
\begin{equation}
\label{eq-rdf}
F(\vec{r},v_{||}; t) = \int \limits_0^{\infty} f(\vec{r},v_{||},v_g^2; t) d v_g^2 = P \frac{v_0^2}{\tau} \int \limits_0^{\infty} \exp\left(-\frac{v_g^2}{\tau(\vec{r},v_{||}; t)}\right) dv_g^2.
\end{equation}

The original kinetic equation \ref{eq-fkin} cannot be exactly integrated in respect of $v_g^2$ in general case because some coefficients depend on the gyro-velocity. The idea is to substitute the gyro-velocity by the "perpendicular  temperature", which is the average gyro-velocity. Using the distribution function form \ref{eq-fmaxw}, it is easy to show that

\begin{equation}
\langle v_g^2 \rangle = \frac {\int \limits_0^{\infty} v_g^2 f(\vec{r},v_{||},v_g^2; t) d v_g^2}{\int \limits_0^{\infty} f(\vec{r},v_{||},v_g^2; t) d v_g^2} = \tau(\vec{r},v_{||}; t).
\end{equation}

\subsection{Integrated kinetic equation}

Here we integrate the kinetic equation, each effect is considered separately.

Firstly, several terms in the equation \ref{eq-fkin} don't depend on $v_g^2$ and their integration is trivial:

\begin{eqnarray}
\int \limits_0^\infty \left( \frac{\partial f}{\partial t} \right)_1 dv_g^2 &+& \int \limits_0^\infty (\vec{u}^* + v_{||}\vec{b}) \frac{\partial f}{\partial \vec{r}} dv_g^2 +%
 \int \limits_0^\infty \frac qm \vec{\mathcal{E}}\cdot \vec{b} \frac{\partial f}{\partial v_{||}} dv_g^2 = \nonumber \\
&& \left(\frac{\partial F}{\partial t}\right)_1 + (\vec{u}^* + v_{||}\vec{b}) \frac{\partial F}{\partial \vec{r}} + \frac qm \vec{\mathcal{E}}\cdot \vec{b} \frac{\partial F}{\partial v_{||}}.
\end{eqnarray}
Next, we use the average squared gyro-velocity for the perpendicular $\nabla B$ drift:

\begin{eqnarray}
\int \limits_0^\infty \left( \frac{\partial f}{\partial t} \right)_2 dv_g^2 &+& \int \limits_0^\infty \vec{u}_{\nabla B} \frac{\partial f}{\partial \vec{r}} dv_g^2 \approx \nonumber \\%
&& \int \limits_0^\infty \left( \frac{\partial f}{\partial t} \right)_2 dv_g^2 + \vec{U}_{\nabla B} \int \limits_0^\infty \frac{\partial f}{\partial \vec{r}} dv_g^2 = %
\left(\frac{\partial F}{\partial t}\right)_2 + \vec{U}_{\nabla B} \frac{\partial F}{\partial \vec{r}},
\end{eqnarray}
where 
\begin{equation}
\vec{U}_{\nabla B} = \frac mq \frac{\vec{B} \times \vec{\nabla}B}{2 B^2} \tau.
\end{equation}
Finally, we use average squared velocity to describe the $v_{||}$ variation due to the magnetic mirroring:

\begin{eqnarray}
\int \limits_0^\infty \left( \frac{\partial f}{\partial t} \right)_3 dv_g^2 &+& \int \limits_0^\infty  \left(- v_g^2 \frac {\vec{\nabla} B \cdot \vec{b}}{2 B}\right) \frac{\partial f}{\partial v_{||}} dv_g^2 \approx \nonumber \\
&& \int \limits_0^\infty \left( \frac{\partial f}{\partial t} \right)_3 dv_g^2 + \int \limits_0^\infty  \left(- \tau \frac {\vec{\nabla} B \cdot \vec{b}}{2 B}\right) \frac{\partial f}{\partial v_{||}} dv_g^2 =
 \left( \frac{\partial F}{\partial t} \right)_3 - \frac 12 \mathcal{G} \tau \frac{\partial F}{\partial v_{||}},
\end{eqnarray}
where
\begin{equation}
\mathcal{G} = \frac {\vec{\nabla} B \cdot \vec{b}}{B}.
\end{equation}
Integrating the last term containing the $\frac{\partial f}{\partial v_g^2}$ in respect of $v_g^2$ is, obviously, zero. Therefore, one can write an approximate drift-kinetic equation based on the $v_g$-averaging:
\mgc{
\begin{equation}
\hat{L}_R F = \frac{\partial F}{\partial t} + (\vec{u}^* + \vec{U}_{\nabla B}+ v_{||}\vec{b}) \frac{\partial F}{\partial \vec{r}} + \frac qm \vec{\mathcal{E}}\cdot \vec{b} \frac{\partial F}{\partial v_{||}} - \frac 12 \mathcal{G} \tau \frac{\partial F}{\partial v_{||}} = 0.
\end{equation}
}

\subsection{Energy equation}

Now we need an equation governing the evolution of the "perpendicular temperature" $\tau$. Since it is assumed that the magnetic moment $v_g^2/B$ is conserved, similar to approximations above, we assume that the "average moment" is conserved as well

\[
\mathcal{M} = \frac {\int \limits_0^{\infty} \frac {v_g^2}{2B} f(\vec{r},v_{||},v_g^2; t) d v_g^2}{\int \limits_0^{\infty} f(\vec{r},v_{||},v_g^2; t) d v_g^2} = \frac{\tau}{2B}.
\]
Mathematically, this can be written as 
\begin{equation}
\hat{L}_R \mathcal{M}F = \frac 12 \hat{L}_R \left( \frac{\tau}{B}F \right) = 0.
\end{equation}
Expanding the above equation yields
\begin{eqnarray}
&&\frac{\partial F}{\partial t}\frac{\tau}{B} + (\vec{u}^* + \vec{U}_{\nabla B}+ v_{||}\vec{b}) \frac{\partial F}{\partial \vec{r}}\frac{\tau}{B} + \frac qm \vec{\mathcal{E}}\cdot \vec{b} \frac{\partial F}{\partial v_{||}}\frac{\tau}{B} - \frac 12 \mathcal{G} \tau \frac{\partial F}{\partial v_{||}}\frac{\tau}{B} + \nonumber \\
&&\;\; \frac{\partial \tau}{\partial t}\frac{F}{B} + (\vec{u}^* + \vec{U}_{\nabla B}+ v_{||}\vec{b}) \frac{\partial \tau}{\partial \vec{r}}\frac{F}{B} - (\vec{u}^* + \vec{U}_{\nabla B}+ v_{||}\vec{b}) \vec{\nabla}B \frac{F \tau}{B^2} + \frac qm \vec{\mathcal{E}}\cdot \vec{b} \frac{\partial \tau}{\partial v_{||}}\frac{F}{B} - \frac 12 \mathcal{G} \tau \frac{\partial \tau}{\partial v_{||}}\frac{F}{B} =0. \nonumber 
\end{eqnarray}
Substracting $\frac{\tau}{B} \hat{L}_RF$ and multiplying by $\frac{B}{F}$ yields the following:
\mgc{
\begin{equation}
\frac{\partial \tau}{\partial t} + (\vec{u}^* + \vec{U}_{\nabla B}+ v_{||}\vec{b}) \frac{\partial \tau}{\partial \vec{r}} - (\vec{u}^* + \vec{U}_{\nabla B}+ v_{||}\vec{b}) \vec{\nabla}B \frac{\tau}{B} + \frac qm \vec{\mathcal{E}}\cdot \vec{b} \frac{\partial \tau}{\partial v_{||}} - \frac 12 \mathcal{G} \tau \frac{\partial \tau}{\partial v_{||}} =0.
\end{equation}
}

\subsection{Field equations}

Electric and magnetic field evolution can be described using the Maxwell equations
\begin{eqnarray}
\frac{\partial \vec{B}}{\partial t} &=& - \vec{\nabla} \times \vec{E} \\
\frac{\partial \vec{E}}{\partial t} &=& \frac{1}{\epsilon_0 \mu_0} \vec{\nabla} \times \vec{B} - \frac 1{\epsilon_0}\vec{j} \\
\vec{\nabla}\cdot \vec{E} &=& \frac{\rho}{\epsilon_0}\\
\vec{\nabla}\cdot \vec{B} &=& 0,
\end{eqnarray}
where
\[
\rho = \Sigma_s q_s \left[ \int \limits_V F(\vec{r},v_{||}; t) dv_{||}\right]
\]
and
\[
\vec{j} = \Sigma_s q_s \left[ \int \limits_V (\vec{u}^* + \vec{U}_{\nabla B}+ v_{||}\vec{b})F(\vec{r},v_{||}; t) dv_{||}\right].
\]

\section{Numerical tests}

Here we investigate the magnetic mirroring, which would produce the highest systematic error in reduced kinetic approximation. The results of simple numerical tests comparing the full kinetic solutions with reduced kinetic solutions in stationary magnetic field are shown below. The magnetic field distribution is shown in Figure 1. The first test (Figure 2) shows the evolution of the distribution functions for particles moving through the weak magnetic mirrors (i.e. most particles are in the loss cone), the second test (Figure3) is for particles oscillating between two strong mirrors (i.e. most particles are outside the loss cone). All test models here are one-dimensional, with particles moving along the axes of cylindrically symmetric magnetic configurations (i.e., $\vec{B} \times \vec{\nabla}B = 0$). 

It can be seen that the reduced kinetic solution is similar to the 'full kinetic' solution, although the former is more compact. This is not surprising, because of the introduced averaging over the gyro-velocity: a single value of $dv_{||}/dt$ results in a lower dispersion in $v_{||}$ and, hence, in $x$.

\clearpage

\begin{figure*}[h!]    %%%%%%%%%%%%%%%%%% FIGURE 1 
\centerline{\includegraphics[width=0.45\textwidth,clip=]{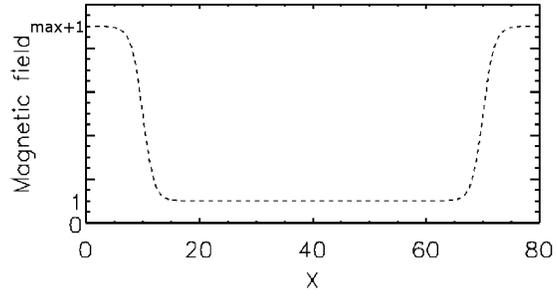}}
\caption{Magnetic field distribution in the 1D test simulations. The value of magnetic field at the boundaries, {\it max}, is equal 1 in the model with weak convergence and 32 in the model with strong convergence.}
\end{figure*}

\begin{figure*}[h!]    %%%%%%%%%%%%%%%%%% FIGURE 1 
\centerline{\includegraphics[width=0.9\textwidth,clip=]{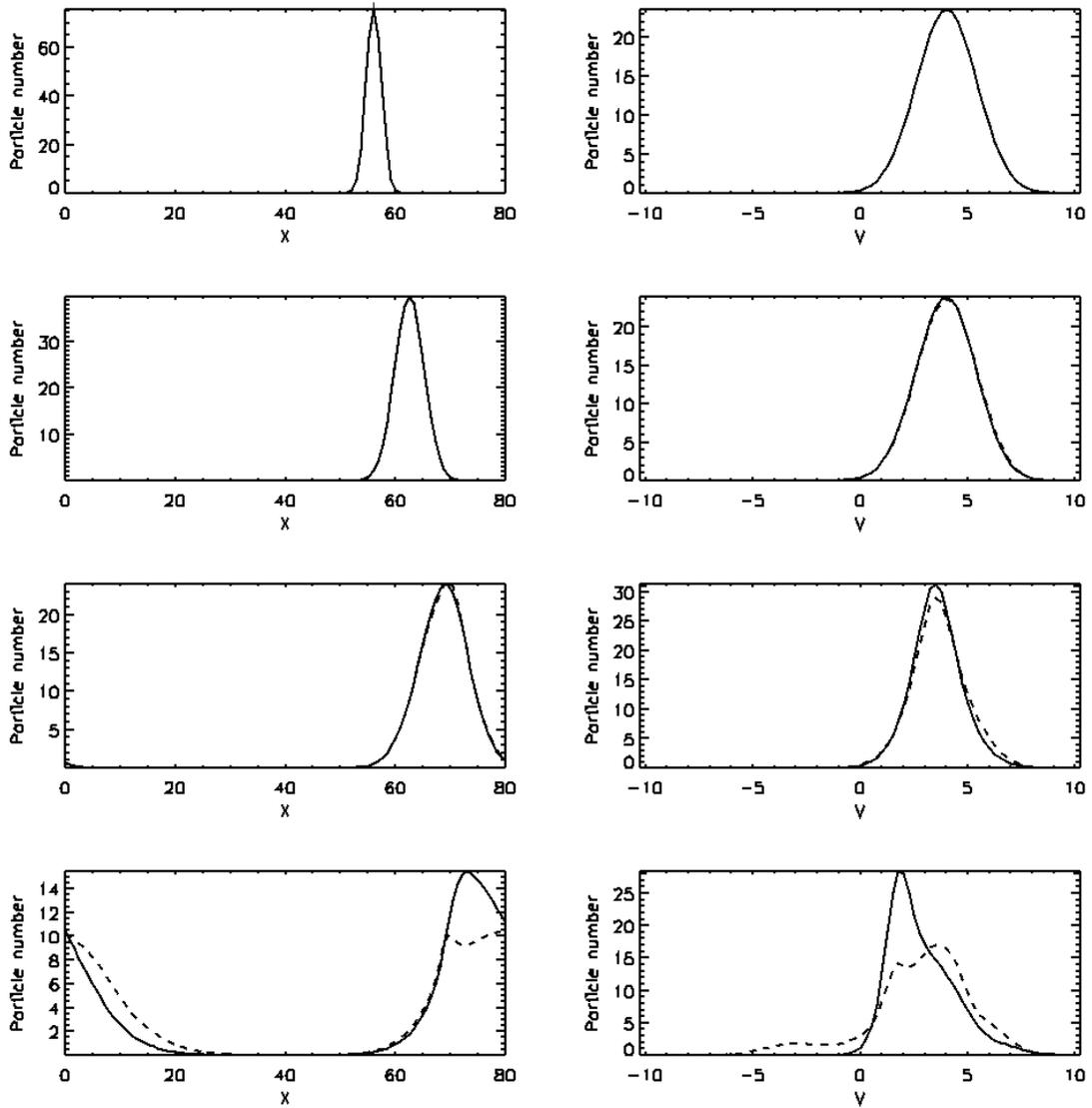}}
\caption{Particle number versus position (left panels) and velocity (right panels). The system is periodic in x. The magnetic mirrors are located at the left and right boundaries. Magnetic convergence ratio is 2, magnetic 'cork' thickness is 16. The initial particle velocity is 4, the initial velocity dispersion is 2, and the initial gyro-temperature is 4 (corresponding to the velocity dispersion of 2). Panels from top to bottom correspond to time 0, 32, 64, 128.}
\end{figure*}

\begin{figure*}[h!]    %%%%%%%%%%%%%%%%%% FIGURE 1 
\centerline{\includegraphics[width=0.9\textwidth,clip=]{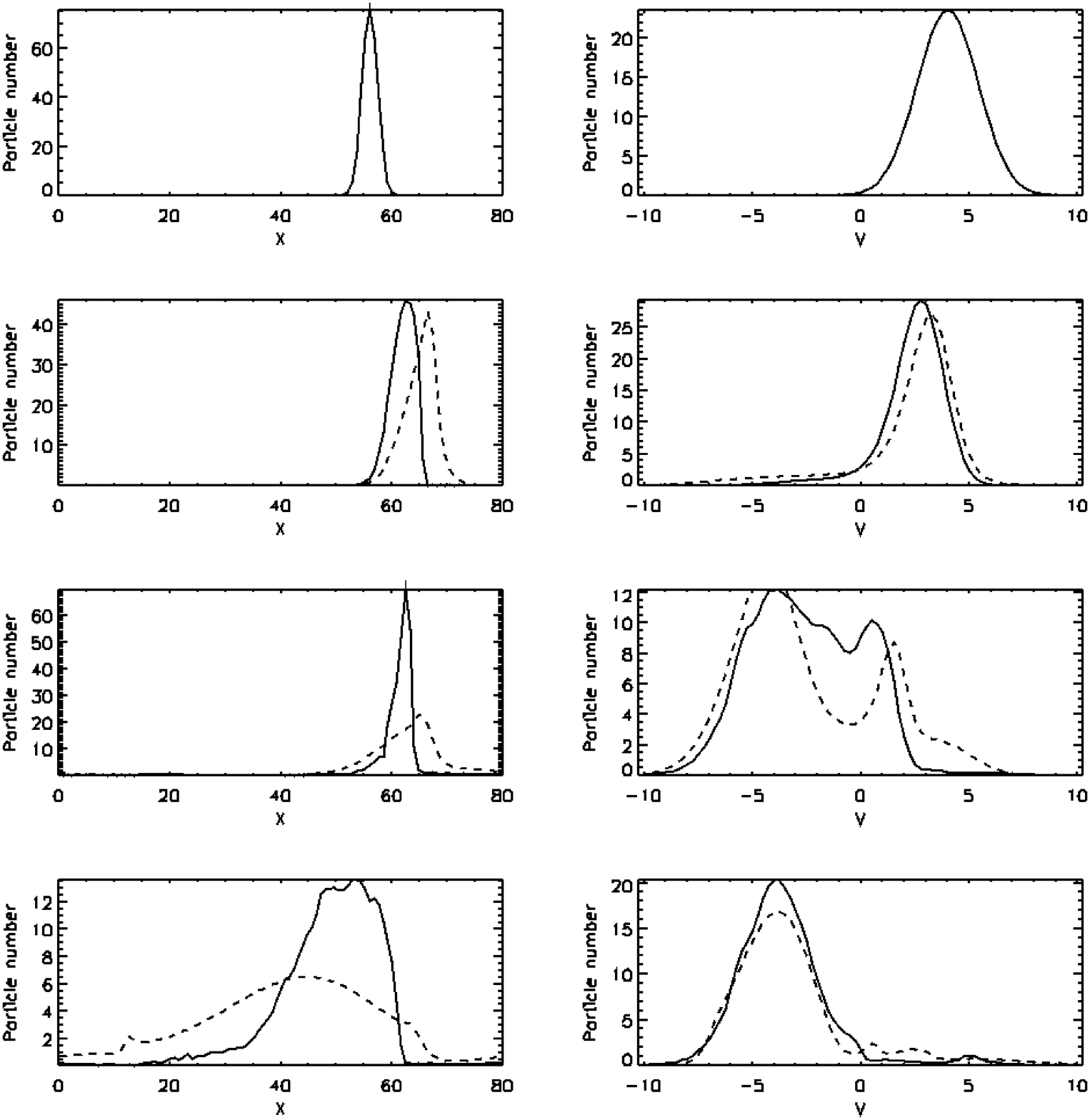}}
\caption{Same as in Figure 1, but the magnetic convergence is 32.}
\end{figure*}

\end{document}